\documentclass[aps]{revtex4}

\usepackage{graphicx}
\usepackage{dcolumn}
\usepackage{bm}
\usepackage{longtable}

\newcommand{\nuc}[2]{$ ^{#1}\mbox{#2}$}
\newcommand{\pbar}{$\bar{\mbox{p}}$}
\newcommand{\nbar}{$\bar{\mbox{n}}$}
\newcommand{\fh}{$f_{\mathrm{halo}}$}
\newcommand{\dr}{$\Delta r_{np}$}
\newcommand{\da}{$\Delta a_{np}$}
\newcommand{\dc}{$\Delta c_{np}$}
\newcommand{\drf}{$\Delta r^f_{np}$}
\newcommand{\daf}{$\Delta a^f_{np}$}
\newcommand{\dcf}{$\Delta c^f_{np}$}
\newcommand{\app}{$a_p$}
\newcommand{\an}{$a_n$}
\newcommand{\cp}{$c_p$}
\newcommand{\cn}{$c_n$}
\newcommand{\rp}{$r_p$}
\newcommand{\rn}{$r_n$}

\newcommand{\gl}{$\Gamma_{low}$}
\newcommand{\gu}{$\Gamma_{up}$}
\newcommand{\pb}{$^{208}\mbox{Pb}$}
\newcommand{\bi}{\nuc{209}{Bi}}

\newcommand{\nn}{$N \overline{N}$}

\newcommand{\ea}{{\it et al.}}

\begin{document}

\pagestyle{myheadings}
\markright{\today}


\title{Neutron density distributions from antiprotonic \nuc{208}{Pb}
and \nuc{209}{Bi} atoms}

\author{B.~K{\l}os}
\affiliation{Physics Department, Silesian University,
                                PL-40-007 Katowice, Poland}
\author{
A.~Trzci{\'n}ska\footnote{e-mail: agniecha@slcj.uw.edu.pl},
J.~Jastrz\c{e}bski,
T.~Czosnyka\footnote{Tomasz Czosnyka passed away on 19 October, 2006},
M.~Kisieli{\'n}ski,
P.~Lubi{\'n}ski\footnote{Present address: N. Copernicus Astronomical Center, 
Pl-00-716 Warsaw, Poland},
P.~Napiorkowski,
L.~Pie{\'n}kowski}
\affiliation{Heavy Ion Laboratory, Warsaw University,
                                PL-02-093 Warsaw, Poland}

\author{F.~J.~Hartmann, B.~Ketzer, P.~Ring, R.~Schmidt, T.~von~Egidy}
\affiliation{Physik-Department, Technische Universit\"at M\"unchen,
                        D-85748 Garching, Germany}

\author{R. Smola\'nczuk, S.~Wycech}
\affiliation{So{\l}tan Institute for Nuclear Studies,
                                PL-00-681 Warsaw, Poland}

\author{K.~Gulda, W.~Kurcewicz}
\affiliation{Institute of Experimental Physics, Warsaw University,
                                PL-00-681 Warsaw, Poland}

\author{E.~Widmann\footnote{Present address: Stefan Meyer Institute for
    Subatomic Physics, Austrian Academy of Sciences, A-1090 Vienna, Austria}}
\affiliation{CERN, CH-1211 Geneva 23, Switzerland}

\author{B. A. Brown}
\affiliation{Department of Physics and Astronomy, and National
Superconducting Cyclotron Laboratory, Michigan State University,
East Lansing, Michigan 48824-1321, USA}

\date{\today}

\begin{abstract}
  The X-ray cascade from antiprotonic atoms was studied for \pb\ and
  \bi.  Widths and shifts of the levels due to the strong interaction
  were determined.  Using modern antiproton-nucleus optical
  potentials the neutron densities in the nuclear periphery were
  deduced. Assuming two parameter Fermi distributions (2pF) describing
  the proton and neutron densities the neutron rms radii were deduced
  for both nuclei.  The difference of neutron and proton rms radii
  \dr\ equal to $0.16 \pm (0.02)_{stat} \pm (0.04)_{syst}$~fm for \pb\
  and $0.14 \pm (0.04)_{stat} \pm (0.04)_{syst}$~fm for \bi\ were
  determined and the assigned systematic errors are discussed. The
  \dr\ values and the deduced shapes of the neutron distributions are
  compared with mean field model calculations.
\end{abstract}


\maketitle

\section{Introduction}

Beginning more than ten years ago, we have performed an experimental
study of the medium-heavy and heavy antiprotonic atoms using the slow 
antiproton beam from the Low Energy Antiproton Ring (LEAR) at CERN.
The main objective of our program was to obtain information on the
neutron distribution at the nuclear periphery and to provide data useful
for deducing the antiproton-nucleus optical potential parameters.

Two experimental methods were employed. First, using the so called
``radiochemical method'' we have investigated~\cite{jas93,lub94,lub98,sch99}
the ratios of peripheral neutron to proton densities at distances around
2.5~fm larger than the nuclear charge half-density radius~\cite{wyc96}.  The
method consisted in measuring the yield of radioactive nuclei having one
proton or one neutron less than the target nucleus, produced after 
antiproton capture, cascade and annihilation in the target antiprotonic atom.
The experiment yielded 19 density ratios (proportional
to the  ``halo factor'' \fh\ defined below), which were
subsequently employed to deduce the shape of the peripheral neutron
distribution.

The second method consisted in measurements of the antiprotonic-atom 
level widths and shifts due to the strong antiproton-nucleus interaction.
These observables are sensitive to the interaction potential which contains,
in its simplest form, a term depending on the sum of the neutron
and proton densities. The level widths and in a number of cases also the level
shifts were measured for 34 antiprotonic atoms (in some cases for different
isotopes of the same element).

The rich harvest of the two methods employed which are sensitive to
the neutron and proton density ratio and the sum of these densities,
has allowed to derive a number of systematic conclusions on the
nuclear periphery properties presented in a series of summary and
analysis publications~\cite{trz01l,trz01,jas04,trz04}.  Moreover, our
data were used to determine the antiproton-nucleus optical model
parameters through global fits of \pbar\ X-rays and halo
factors~\cite{fri05,wyc06} with a substantially larger and more
precise database than employed in previous
approaches~\cite{bat95,bat97}.

Besides our summary papers, after the end of the antiprotonic X-ray (PS209)
experiment, we prepared more detailed reports, containing information
on experimental procedures and their analysis for some cases
studied~\cite{sch98,har01,sch03,klo04}. The present article, 
dealing with \nuc{208}{Pb} and \nuc{209}{Bi} antiprotonic atoms,
is the next in this
series.

During the last years it was shown that properties of the
\nuc{208}{Pb} neutron distribution can be correlated with a number of
quantities in various fields.  In particular the knowledge of the
difference \dr\ between the rms radii of neutrons and protons in this
nucleus constrains the symmetry energy of nuclear matter and therefore
is reflected in the neutron Equation of State
(EOS)~\cite{bro00,fur02,die03,ste05,yos06,ste05b}. The neutron EOS
models, in turn, are used to calculate the properties of neutron
stars, such as their radii and proton
fraction~\cite{hor01a,hor01b}. However, not only the first moments of
the neutron density distributions, but also their shapes are of
considerable interest, e.g. in the determination of the isovector
potential parameter of the pion-nucleus \mbox{s-wave} interaction in
nuclear matter~\cite{gei02} or in the calculation of the lepton flavor
violating muon-electron conversion rate~\cite{kit02}.  There is also a
certain dependence on the radial neutron distribution in the proposed
determination of the \pb\ neutron rms radius through parity violating
electron-nucleus scattering~\cite{hor01c}.

Experimentally, the \dr\ value in \nuc{208}{Pb} was previously determined
using hadronic probes (elastic scattering, inelastic scattering exciting GDR),
reported in Refs.~\cite{ray79,hof80,kra94,kra99,sta94,csa03,kra04,kar02} and discussed
in~\cite{jas04}. 
There were also some 
 attempts to deduce higher moments of the neutron distribution from the
hadron scattering experiments~\cite{mac95} (see also~\cite{kar02}). 
On the other hand, the ability of the medium-energy elastic
proton scattering data to determine the neutron distributions was recently
contested~\cite{pie06}.

The measurement in \bi\ offers other advantages. It is the only
experiment that allows to see an even-odd isotopic effect in heavier
nuclei, in this case due to the loosely bound proton in \bi.  One
difficulty in the way of analysis is the calculation of the
hyperfine-structure that is comparable to the strong interaction shift
and broadening. After this is done, it turns out that the level shift
in \pb\ is repulsive (as most of the lower shifts), but the level
shift in \bi\ is attractive.  This finding is open to
interpretation. Here we pursue the view that it is related to a \pbar
N quasi-bound state, which is important in cases of loosely bound
valence nucleons.

\section{Experimental methods}

The heavy antiprotonic atoms $^{208}$Pb and $^{209}$Bi were investigated
during the experiment PS209 at CERN in 1996 using antiprotons of momentum 
106~MeV/c. 
Table~\ref{target} gives the target properties and the number of antiprotons
used for each target.

The antiprotonic X-rays emitted during the antiproton cascade were measured by
three high-purity germanium (HPGe) detectors. Two detectors were coaxial with
an active diameter of 49\,mm and a length of 50\,mm (relative efficiency about
19\% and 17\%, respectively),
and the third one was planar with 36~mm diameter and
a thickness of 14~mm.  The detectors were placed at distances of about 50~cm from
the target at angles of 13$^\circ$, 35$^\circ$ and 49$^\circ$ 
towards the beam axis,
respectively. The detector-target distance was adjusted to obtain a good
signal-to-noise ratio and   
to decrease at the same time the background produced by
pions from the annihilation processes.

More details concerning the experimental methods, the detector calibration
and the data reduction may be found in our previous
publications~\cite{sch98,har01,sch03,klo04}.

\section{Experimental results}

\begin{figure}[h]
\includegraphics[width=0.5\textwidth]{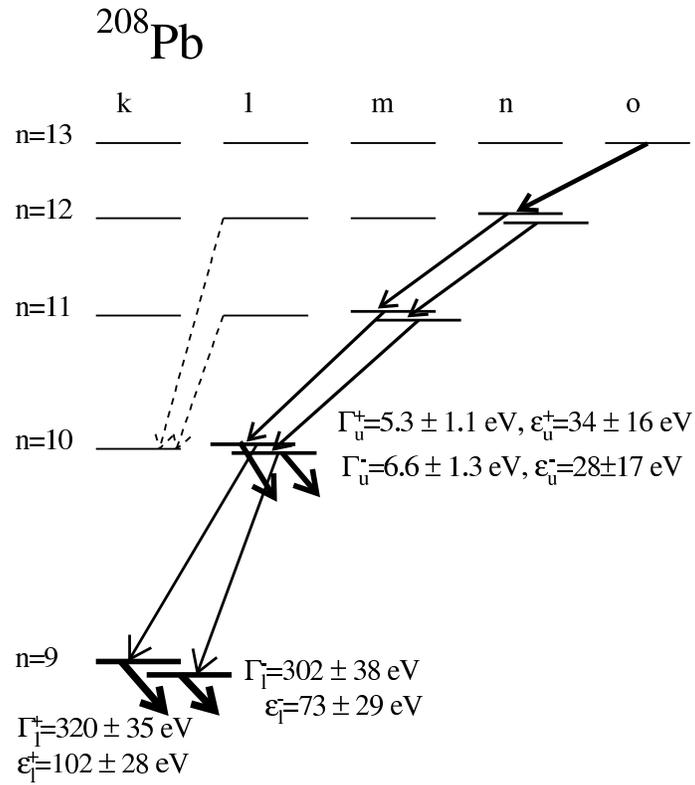}
\caption{\label{observ} Summary of shifts and widths measured for \pb.}
\end{figure}

The strong interaction between antiproton and nucleus causes a
sizeable change of the energy of the last X-ray transition from its
purely electromagnetic value. The nuclear absorption reduces the
lifetime of the lowest accessible atomic state (the ``lower level'',
which for lead is the $(n,l=9,8)$ state) and hence this X-ray line is
broadened.  Nuclear absorption also occurs from the next higher level
(``upper level'') although the effect on level energy and width is
generally too small to be directly measured. The width of the
$(n,l=10,9)$ level was deduced indirectly by measuring the 
intensity loss of the final X-ray transitions. The level scheme for the
antiprotonic Pb atom with the observables of the X-ray
experiment is shown in Fig.~\ref{observ}.

\begin{figure}[h]
\includegraphics[width=0.75\textwidth]{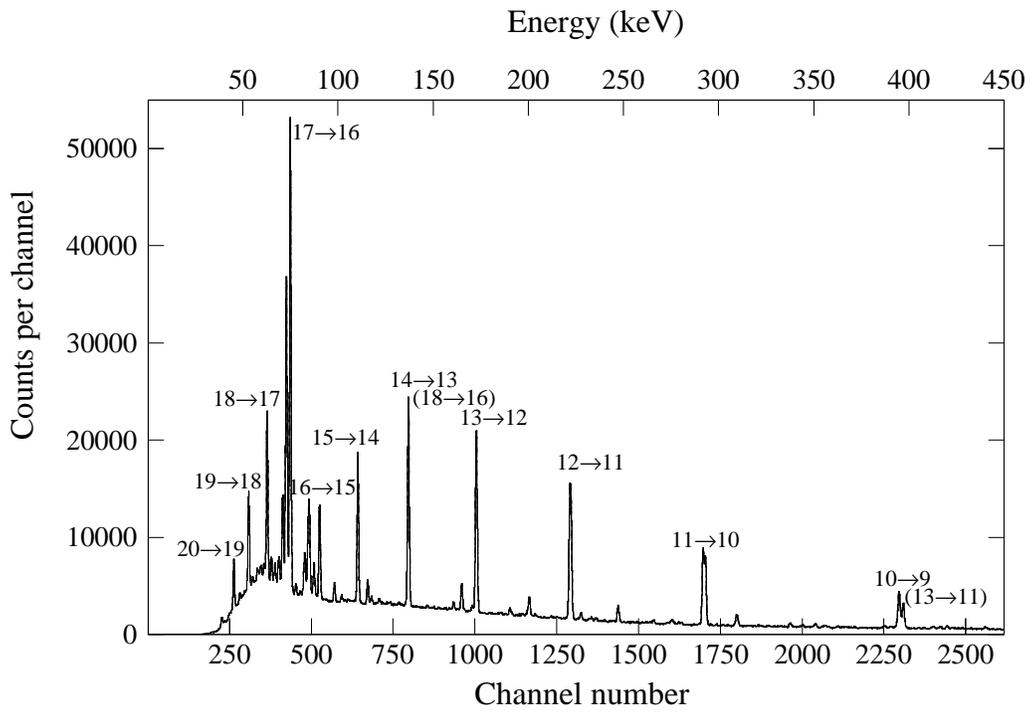}
\caption{\label{spect} Antiprotonic X-ray spectrum from $^{208}$Pb 
measured with the  HPGe detector of 19\% relative efficiency.}
\end{figure}

\begin{figure}[h]
\includegraphics[width=0.75\textwidth]{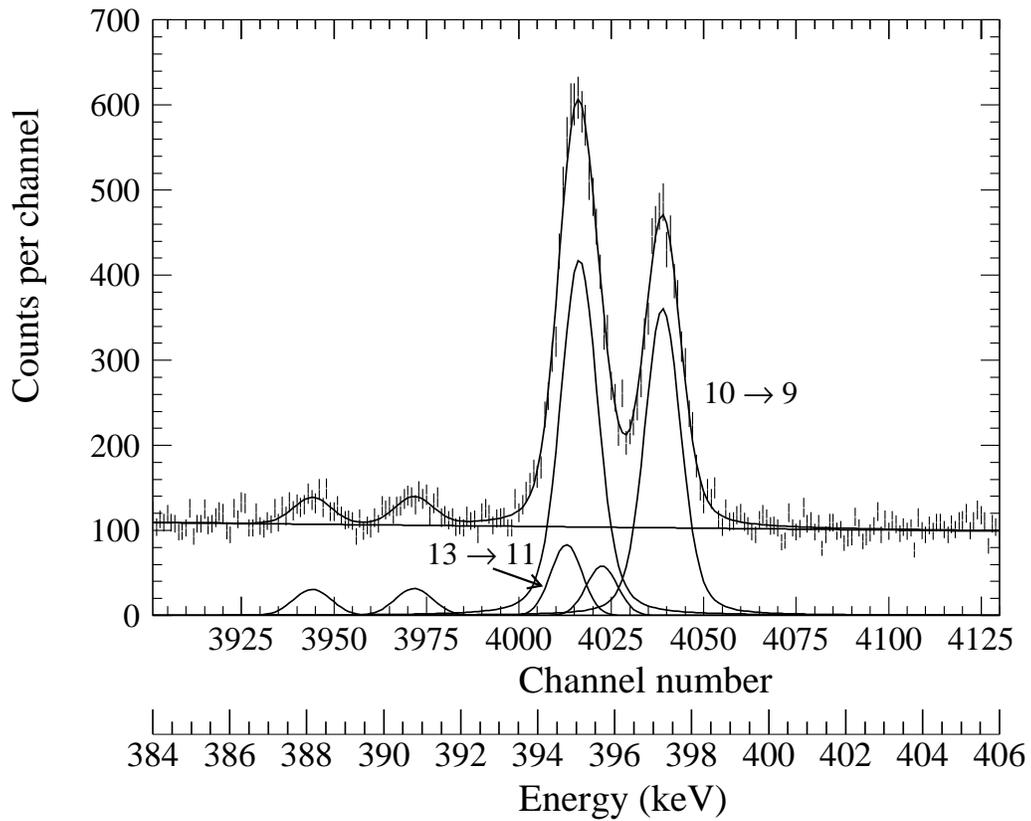}
\caption{\label{voigt} Part of the antiprotonic X-ray spectrum 
measured for  $^{208}$Pb using
  the detector with the 1035\,mm$^2$ x 14\,mm crystal. The fit to the broadened
  10$\to$9 transition is also shown. The 13$\to$11 line is admixed to the
  10$\to$9 line.}
\end{figure}

The X-ray spectrum measured with antiprotons stopped in $^{208}$Pb is
shown in Fig.~\ref{spect}.  Those lines in the spectra that are not
broadened were fitted with Gaussian profiles. The lowest observable
LS-split doublet lines $(n=10\to9)$, which are significantly
broadened,\ were fitted with two Lorentzians convoluted with Gaussians
(Fig.~\ref{voigt}). 

\begin{figure}[h]
\includegraphics[width=0.75\textwidth]{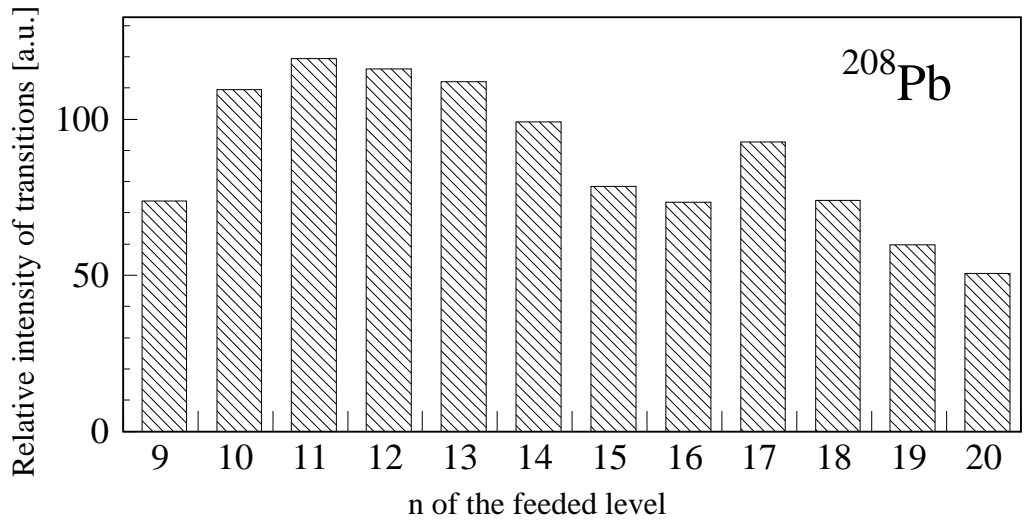}
\caption{\label{his} Total relative intensities of the observed
  transitions feeding the indicated n-level in \pb\ normalized to the
  transition n=13$\to$12, taken as 100. }
\end{figure}

The measured relative intensities of the antiprotonic X-rays observed in the
investigated lead and bismuth targets are given in Table~\ref{relat}. These
intensities were used to determine the feeding of the consecutive $n$ levels
along the antiprotonic-atom cascade. This is shown for \pb\ in Fig.~\ref{his}.

\begin{figure}[h]
\includegraphics[width=0.9\textwidth]{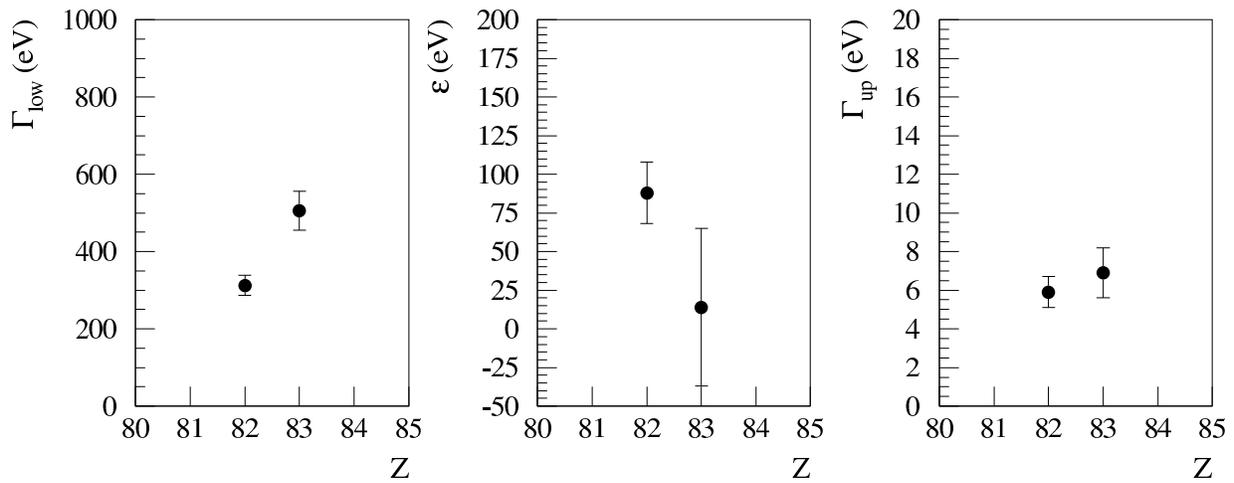}
\caption{\label{isoef} Width and shift of the levels $(n,l=9,8)$ and
  widths of the levels $(n,l=10,9)$ plotted versus $Z$, for
  \nuc{208}{Pb} and \nuc{209}{Bi}. Positive level shifts correspond to
  repulsive interactions. The presented lower level shift and width
  for \bi\ are not corrected for the hyperfine contribution.  }
\end{figure}

Table \ref{shiftt} gives the measured shifts $\varepsilon$, defined by
$\varepsilon=E_{\mathrm{em}}-E_{\mathrm{exp}}$, where $E_{\mathrm{exp}}$ 
the experimental value for the transition energy and $E_{\mathrm{em}}$ the
energy calculated without strong interaction~\cite{bor83}. For the lead
$(n,l=9,8)$ levels the shifts are clearly repulsive, whereas for bismuth
the levels shifts are consistent, within the errors, with zero
(Fig.~\ref{isoef}).  However, as discussed in the next sections, there exists an
additional repulsive shift due to the hyper fine structure. For the
$(n,l=10,9)$ levels in lead the shifts are repulsive, as 
in the case of the $(n,l=9,8)$
levels, but the shifts are smaller. 
Tables \ref{gamlt} and
\ref{gamut} 
give the measured widths. As indicated above, the
widths of the $(n,l=10,9)$ levels were derived from the intensity balance of
transitions feeding and depopulating these levels.  Contributions of parallel
transitions to the measured intensities were obtained from cascade
calculations (see~\cite{sch98}). The rates for radiative dipole transitions
were calculated with the formulae given in Ref.~\cite{eis61}. The Auger rates
were derived from the radiative rates and from cross sections for 
photoeffect using Ferrell's formula~\cite{fer60}.  The width of the levels
$(n,l=10,9)$ are larger for \bi\ than for \pb\ 
(Fig.~\ref{isoef}).  This is due to the hyperfine structure in Bi, which
will be discussed below.  

\section{Analysis and discussion of the \nuc{208}{Pb} measurement}

  \subsection{Introductory statements}
  \label{intro2}

The analysis of the presented antiprotonic \nuc{208}{Pb} atom
data is based on some assumptions that are briefly 
mentioned here and discussed below.
We assume 
and show by a comparison with model calculations
that charge, proton and neutron distributions in this
nucleus can be well approximated by two-parameter
Fermi (2pF) distributions:
$\rho(r) = \rho_0 \cdot \{1+\mathrm{exp}(\frac{r-c}{a})\}^{-1}$, where
$c$ is the half-density radius, $a$ is the diffuseness parameter, 
and $\rho_0$ is a normalization factor. 
In particular, calculating the neutron
rms radius from the antiprotonic X-ray data sensitive to densities
at distances around 1.5~fm larger than the half-density charge
radius~\cite{wyc96}, we extrapolate the experimental density well
into the interior of the nucleus.
As shown below (Sec.~\ref{calc-mean}), this assumption is in reasonable
agreement with the density shapes calculated in terms of the mean field
models.

In evaluating the observables of the antiproton-nucleus interaction
the important question of the ratio of the annihilation probability on
a neutron to the one on a proton arises. In the simplest nuclear
optical potentials this ratio is given by the ratio of the imaginary
parts of the effective scattering lengths, $R =Im \,b_0^{\; n} / Im \,
b_0^{\; p}$.  The experimental determination of this
quantity~\cite{bug73,wad76} gave $R=0.63$.

In spite of this observation, a value R=1 was assumed in the optical
potentials proposed in Refs.~\cite{bat95,bat97,fri05}.  Our analysis
of the antiprotonic X-ray data, together with the radiochemical
experiment, also indicated~\cite{trz01th} a better consistency of
these two methods when $R=1$ was chosen. Therefore this value is also
adopted in the present data evaluation, as discussed in the following
sections.

 \subsection{Charge and proton distributions}
  
\begin{figure}[h]
\includegraphics[width=0.5\textwidth]{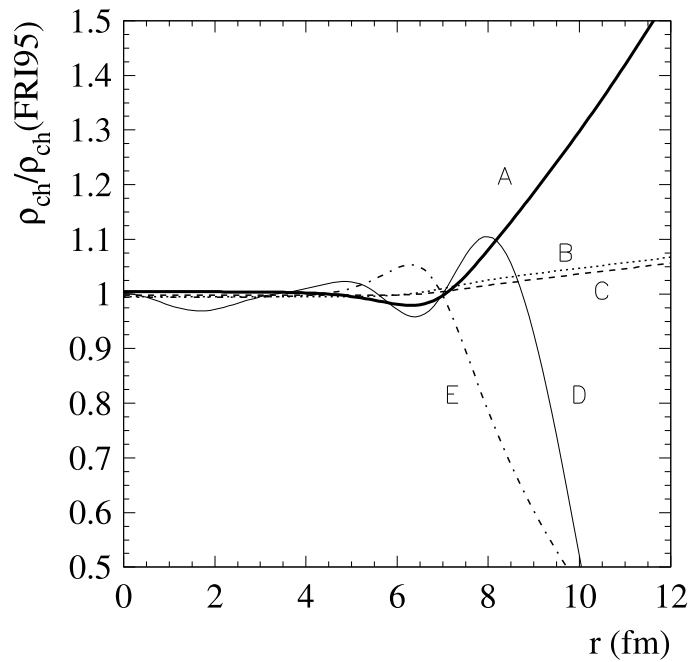}
\caption{\label{pbcharge}Comparison of \nuc{208}{Pb} charge-density
  distributions deduced from various compilations. The plotted charge
  density distributions are normalized to the one given by
  Fricke~\cite{fri95}. Other charge distributions (charge rms radii in
  parenthesis) are taken from: (A)~\cite{jag74} (5.521~fm),
  (B)~\cite{jen71} (5.515~fm), (C)~\cite{kes75} (5.510~fm),
  (D)~\cite{vri87} (5.503~fm), (E)~\cite{nif69} (5.46~fm). The charge
  rms radius given in~\cite{fri95} is 5.504~fm.}
\end{figure}

  It is generally assumed (see e.g. Ref.~\cite{fri95}) that the charge
  rms values are known for the stable nuclei with remarkable
  precision, about 0.3\%. The same belief is often projected on the
  charge {\it distributions}.  In Fig.~\ref{pbcharge} we show (as it
  was already observed in~\cite{klo04}) that this is not the case for
  the radial distances where the antiproton-nucleus interaction takes
  place in \nuc{208}{Pb} (about 7 to 10~fm away from the nuclear
  center, see below).  In this figure the charge density of
  \nuc{208}{Pb}, tabulated in a number of compilations, is compared
  with the most recent one by Fricke {\it et al.}~\cite{fri95}.
  Neglecting the oldest tabulation, differences of up to 50\% are
  observed between Fricke \ea, Jager {\it et al.}~\cite{jag74} and de
  Vries {\it et al.}~\cite{vri87} for radial distances close to 10~fm.
  We consequently use the Fricke charge distribution in this work.

  Experiments using electromagnetically interacting probes give charge
  density distributions or rms charge radius values
  (e.g.~\cite{fri95,vri87}) whereas point proton distributions are
  needed when Batty's zero-range antiproton-nucleus optical
  potential~\cite{bat95} is used for the analysis of the experimental
  data.  For the finite-range version of the \pbar-nucleus
  potential~\cite{fri05,wyc06} these point distributions are folded
  over an interaction range.

  In Ref.~\cite{ose90} the analytical formulae to transform the 2pF
  charge distribution to the 2pF point distributions of the proton
  centers were presented.  We have previously used them in our data
  analysis, presented e.g. in Ref.~\cite{trz01}.  Similar analytical
  formulae were recently given in Ref.~\cite{pat03}. In order to
  transform the \nuc{208}{Pb} proton charge distribution of
  Ref.~\cite{fri95}, we have used the proton charge rms radius
  $\sqrt{<r_p^2>} = 0.875$~fm~\cite{codata}, obtaining 2pF point
  proton parameters of $c_p = 6.684$~fm, $a_p = 0.446$~fm and rms $r_p
  = 5.436$~fm.

   \subsection{Calculated mean field neutron and proton distributions}
   \label{calc-mean}
   
   The proton and neutron distributions in the doubly magic \pb\
   nucleus were subject of a large number of theoretical
   investigations. In this paper we select Skyrme-Hartree-Fock (HF)
   and the Hartree-Fock-Bogoliubov models, namely those with the SkP
   (HFB)~\cite{dob84} and SkX (HF) ~\cite{bro98} parametrization, both
   reproducing the \nuc{208}{Pb} charge (proton) radius and neutron
   binding energy remarkably well. (It has, however, recently been
   shown that the SkP Skyrme model may diverge for some nuclei if
   calculated to sufficient accuracy~\cite{len06}).  The third
   self-consistent mean filed model considered here belongs to the
   framework of the relativistic mean field theory (RMF) with the
   recent DD-ME2 parametrization of the effective
   interaction~\cite{lal05}. Although in fitting the DD-ME2 parameters
   the \pb\ \dr\ value (of 0.20~fm) was used to adjust the interaction
   parameters, the {\it shape} of the neutron distribution was
   obtained from the calculation.

\begin{figure}[h]
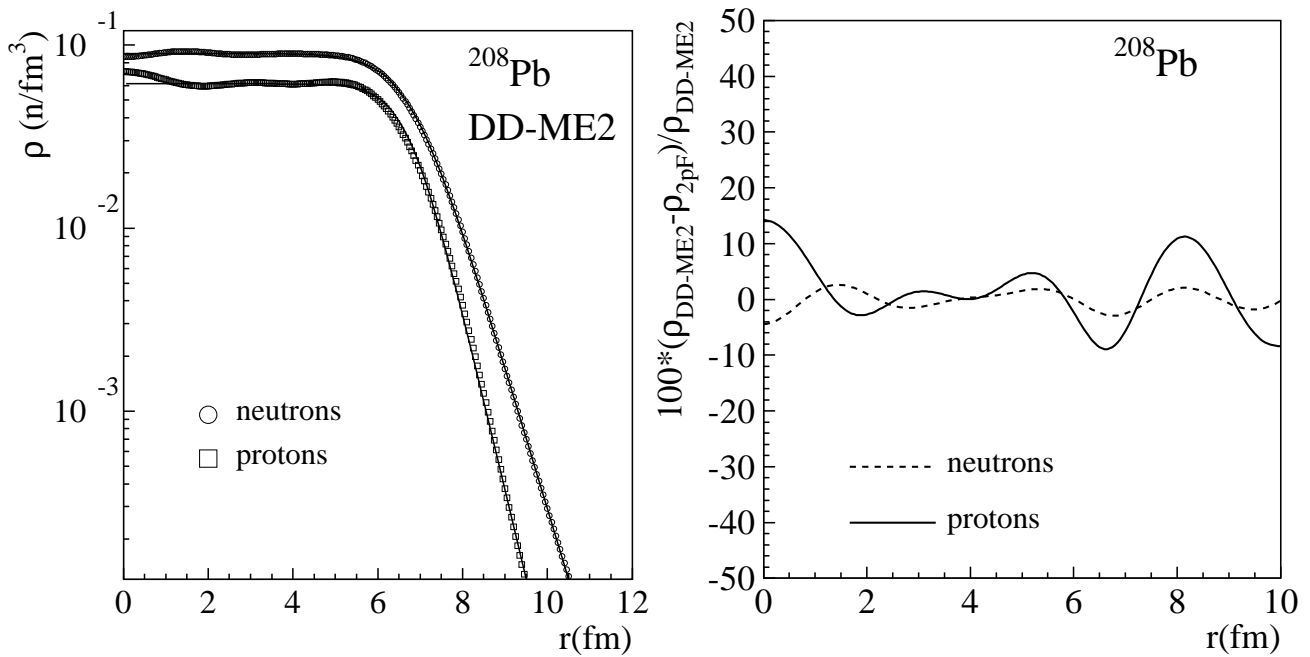

\includegraphics[width=0.475\textwidth]{fig7_fit_ddme2.epsi}
\includegraphics[width=0.475\textwidth]{fig7_dev_2pf_ddme2.epsi}
\caption{\label{ddme22pf} Left panel: two-parameter Fermi (2pF)
  distribution fitted in the range 1-10~fm to proton and neutron
  distributions calculated using the relativistic mean-field theory
  (RMF) with DD-ME2 parametrization~\cite{lal05}. Points -- calculated
  distributions, continuous line -- 2pF fit. Right panel: relative
  differences of fitted (2pF) and calculated (DD-ME2)
  densities. Continuous line -- protons, dashed line -- neutrons.}
\end{figure}

\begin{figure}[h]
\includegraphics[width=0.5\textwidth]{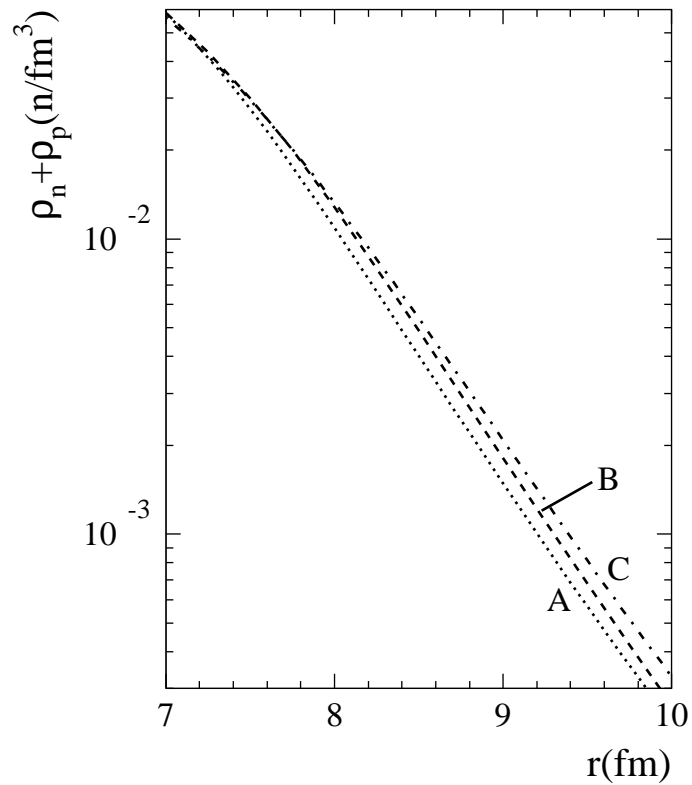}
\caption{\label{nplusp} Sum of the theoretical neutron and proton
  densities for the radial distances at which the antiproton
  annihilation in \pb\ nucleus is significant. (A) -- HF model with
  SkX parameters, (B) -- HFB model with SkP parameters, (C) -- RMF
  model with DD-ME2 parameters.  }
\end{figure}

   Figure~\ref{ddme22pf}, left panel, presents the proton and neutron
   distributions of \pb\ as calculated using the DD-ME2
   parametrization.  This and two other distributions were
   approximated by 2pF distributions fitted to the theoretical
   densities.  Satisfactory fits were achieved for radii between 1 and
   10~fm (local differences between theoretical and fitted densities
   were less than 10\% for protons and less than 4\% for neutrons in
   the radius range 2--9~fm, Fig.~\ref{ddme22pf}, right panel).
   Figure~\ref{nplusp} shows the summed neutron and proton densities
   for the three forces considered.  The relationship between the
   neutron densities, rms radius and equation of state is being
   investigated with a larger set of mean-field models
   in~\cite{bro07}.

\begin{figure}[h]
\includegraphics[width=0.9\textwidth]{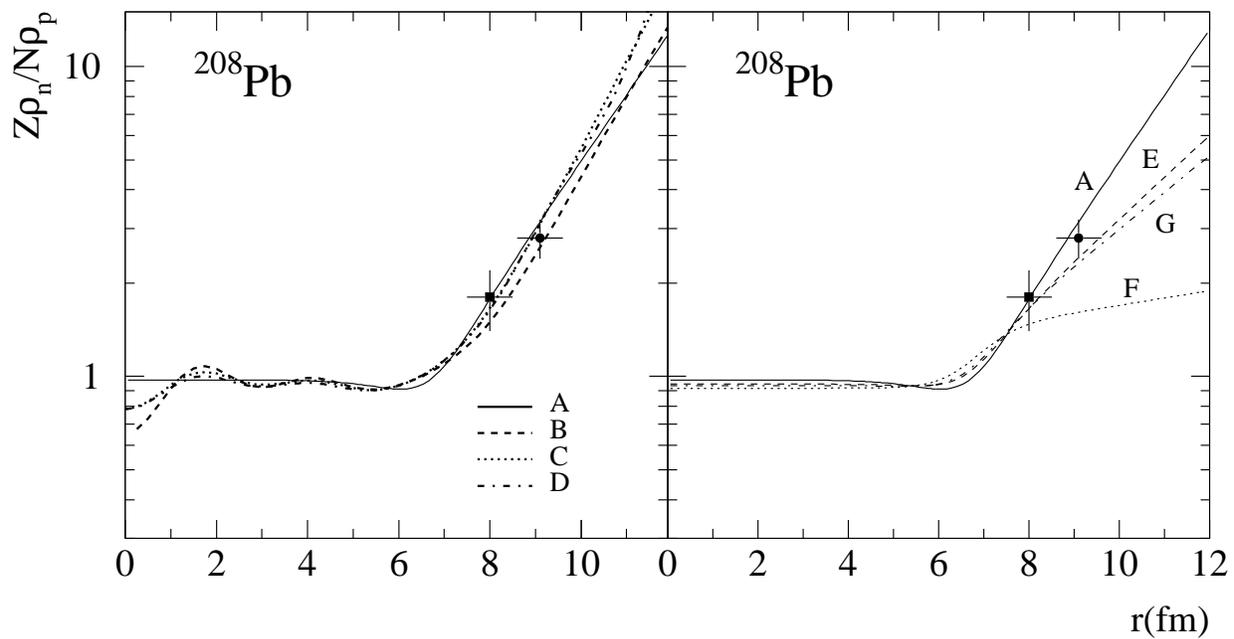}
\caption{\label{rorat-halof} Normalized neutron to proton
  point-nucleon density ratios as a function of the distance from the
  nuclear center in \pb. Left panel (cf. Section~\ref{calc-mean}): (A)
  -- two-parameter Fermi distribution with \dr=0.16~fm and \dc=0, (B)
  -- SkP parametrization, (C) - SkX parametrization, (D) -- DD-ME2
  parametrization. Right panel (cf. Section~\ref{analys}): density
  ratios deduced from two-parameter Fermi distributions: (A) -- the
  same as on the left panel, (E) -- \dr=0.16~fm, \dc=0.1~fm, (F) --
  \dr=0.16~fm, \dc=0.2~fm, (G) \dr=0.17~fm, \dc=0.13~fm
  (cf. Sec.~\ref{wycech}).  The cross at 8~fm represents the halo
  factor from Bugg's experiment~\cite{bug73}, the cross at 9.1~fm is
  the interpolated halo factor (see Sec.~\ref{interpol-halo}).}
\end{figure}

   Table~\ref{2pfpb} gives the results of the fitting procedure. The
   rms radii calculated using $a$ and $c$ values of the 2pF are close
   to those obtained using the theoretical distributions directly.
   The neutron distributions are close to the ``halo
   type''~\cite{trz01}, with $\Delta c_{np} = 0.02$~fm for SkP,
   $\Delta c_{np} = 0.07$~fm for SkX and \dc=0.05~fm for DD-ME2,
   respectively.  This is illustrated in the left-hand part of
   ~Fig.~\ref{rorat-halof}, showing the normalized neutron to proton
   density ratio obtained from the density distributions of the
   discussed models.  The figure presents also the ``pure halo''
   distribution with \dr=0.16~fm and \dc=0~fm.

   \subsection{\label{potential} Antiproton-nucleus optical potentials}
   
   The standard potential for hadronic atoms \cite{bat97} are composed of
   two terms
   \begin{equation}
     \label{O1} V^{opt} = V_S(r) +  {\bf \nabla} V_P(r) {\bf \nabla}
   \end{equation}
   involving proton and neutron components. Both terms, the local $V_S$ and the
   gradient $V_P$ are expected to have the folded form
   \begin{equation}
     \label{O1S}
     V_{S,P} (r) = \frac{2\pi}{\mu_{N \overline{N}}}  
     b_ {S,P}\int d{\bf u} f({\bf u})\rho ({\bf r}-{\bf u}),
   \end{equation}
   which involves nucleon densities $\rho$, folded with a (usually
   Gaussian) function $f$ of some rms radius ,$r_{range}$, of the
   range, "effective lengths " $b$, and some weak dependence of the
   reduced mass ${\mu_{N \overline{N}}}$ on nuclear recoil.  It turned
   out already in a first analysis for heavy atoms \cite{bat97} that
   an independent determination of $ b_ {S}$ and $ b_ {P}$ is not
   realistic.  A simplified result in Ref.~\cite{bat97} gave $b_S = -
   2.5(3) - i \, 3.4(0.3)$~fm with zero $r_{range}$, which we call the
   Batty potential and use in our calculations.  The recent
   phenomenological best fit (Friedman potential) is obtained with a
   single term of $b_{S}\, (=b_0) =-1.3(1) -i \, 1.9(1)$~fm for both
   neutrons and protons and $r_{range}=1.04$~fm~\cite{fri05}.  Within
   all these calculations the X-ray data suggested no significant
   differences in the the values of the p\pbar\ and n\pbar\
   lengths.  No relation of the phenomenological $b_{S}$ to the
   $N\bar{N}$ scattering parameters has been established. Another
   recent analysis \cite{wyc06} attempts to find such a relation from
   the analysis of the lightest atoms H, D, He and of scattering data
   described in terms of the Paris $N\bar{N}$ potential. One important
   difference arises in equation~(\ref{O1S}), which also contains
   nucleon recoil terms not required in a phenomenological approach.
   The nucleon recoil term constitutes about a quarter of the total
   potential and depends on the state of the nucleon. This leads to a
   more complicated parametrization, which will not be repeated here.
   In this potential~\cite{wyc06} one obtains roughly $b_{S}=-1.7 -i
   \, 0.9$~fm, $b_{P}=0 -i \, 0.4$~fm$^3$ with $r_{range}=0.8$~fm and
   the absorptive part of these parameters compares well with the
   p\pbar\ and p\nbar\ scattering data.

   Presently available optical potentials are unable to reproduce the
   level shifts in Pb. This reflects a more general difficulty related
   to uncertainties of the real part of $V^{opt}$. 
   More specific difficulties such as
   \pbar N-quasi-bound states or long-range pion exchange forces in
   the \pbar N system have already been discussed in
   Refs.~\cite{wyc01,wyc06}.

  \subsection{\label{annihil-probab} Annihilation probability}

  The probability $P_{n,l}(r)$ of the nuclear capture from a given
  atomic state $(n,l)$ is defined (for local optical potentials) as $
  P_{n,l}(r) = \mid\phi_{n,l}(r)\mid^2 (Im \, V^{opt}) \cdot r^2$,
  where $\phi_{n,l}$ is the antiprotonic wave function and $r$ is the
  radial distance from the nuclear center.

\begin{figure}[h]
\includegraphics[width=0.6\textwidth]{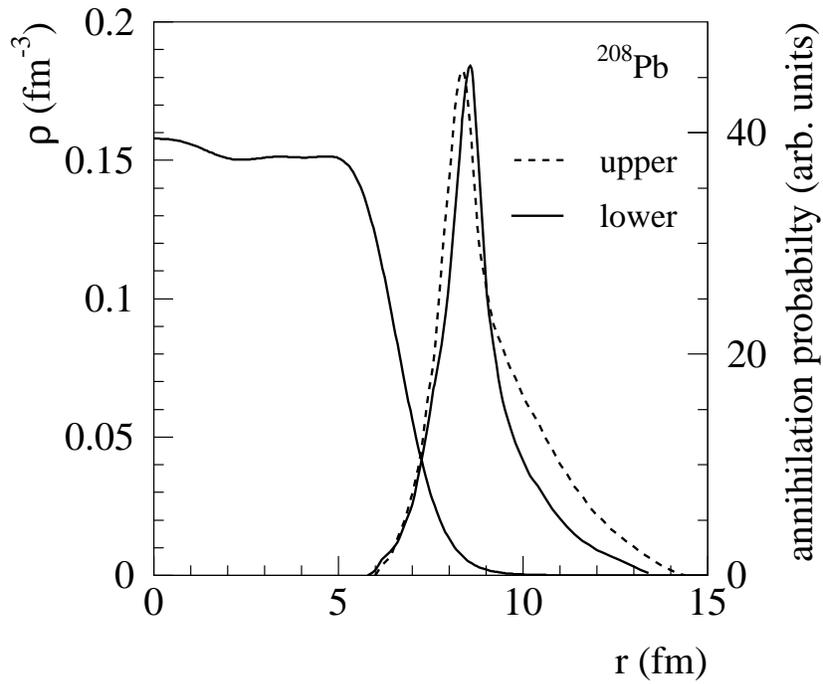}
\caption{\label{annihil-probab-fig} The \pbar\ annihilation
  probability (arbitrary units) from the upper state (dashed line) and
  the lower state (continuous line) for the antiprotonic \pb\ atom. The
  matter density (DD-ME2) in \pb\ is also shown.  The calculations of
  the annihilation probability were done with the Batty zero-range
  potential and 2pF parametrization of the DD-ME2 density
  (cf. Table~\ref{2pfpb}).  }
\end{figure}

  The most probable value of this probability was calculated for some
  cases in Ref.~\cite{wyc96} and the results of these calculations
  were confirmed by the analysis given in Ref.~\cite{fri05}. The
  calculations of the $P_{n,l}(r)$ distributions for \pb\ were
  preformed using Batty potential and density distributions discussed
  in Sec.~\ref{calc-mean}. Figure~\ref{annihil-probab-fig} shows this
  annihilation probability distributions for DD-ME2 densities and
  Table~\ref{annihil-probab-tab} gives parameters of this
  distribution.  Wider distributions (by about 60\% for upper level
  and 20\% for lower level) with more pronounced tails for larger radii
  are obtained using finite range optical potentials.

  \subsection{\label{interpol-halo} Interpolated halo factor}

  In our previous analysis of the antiprotonic X-ray and radiochemical
  data the crucial information used in deducing the shape of the
  neutron distribution (``halo type'' or ``skin type'',
  see~\cite{trz01} for the definition) was the experimentally
  determined halo factor $f_{\mathrm{halo}} = \frac{Y(N_t - 1)}{Y(Z_t
  -1)}\cdot \frac{Z_t}{N_t} \cdot R$; here $Y$ are the yields for the
  $A_t -1$ nuclei, $Z_t$, $N_t$ and $A_t$ are the target proton,
  neutron and mass numbers, respectively, and $R$ was defined and
  discussed above.  We have shown previously that in a number of cases
  for which this halo factor could be measured, it indicated that the
  corresponding 2pF neutron distribution is close to the ``halo
  type'', i.e. with equal proton and neutron radius parameter
  (\cn=\cp) and larger diffuseness parameter for neutrons
  (\an$>$\app). Again, this observation is in fair agreement with the
  mean-field calculations of the nuclear densities (see
  Sec.~\ref{calc-mean}).

\begin{figure}[h]
\includegraphics[width=0.5\textwidth]{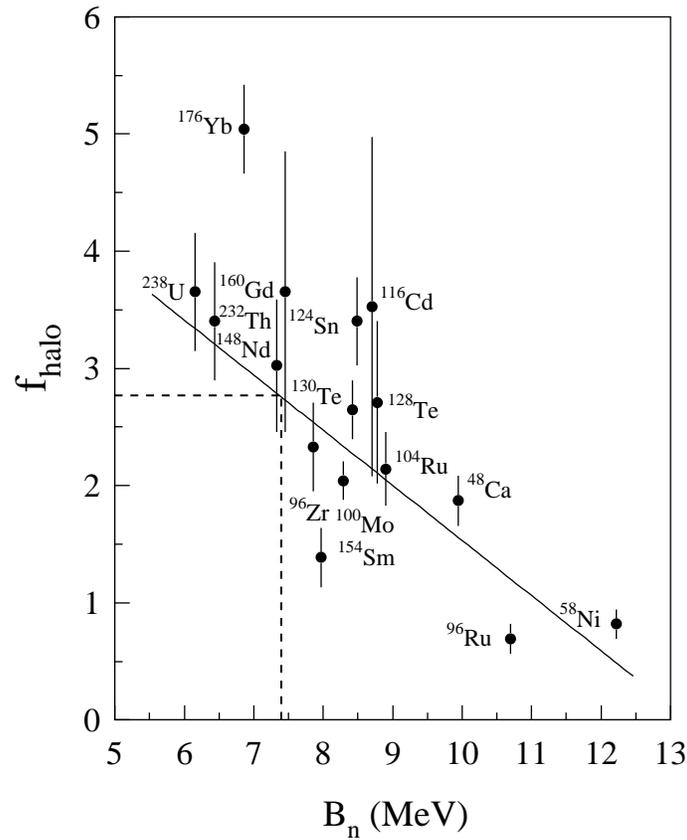}
\caption{\label{pbhalo} A straight line fitted to the experimentally
  determined halo factors~\cite{lub98,sch99,lubth}, plotted as a
  function of the target-nucleus neutron binding energy. For
  \nuc{208}{Pb} the neutron binding energy is equal to 7.4~MeV and the
  interpolated halo factor is $2.8 \pm 0.4$. }
\end{figure}

  In the case of the \nuc{208}{Pb} nucleus, the experimental
  determination of the halo factor by the radiochemical method was not
  possible as one of the ($A_t -1$) isotopes (\nuc{207}{Pb}) is not
  radioactive.  In order to have at least some indication of its value
  the \nuc{208}{Pb} halo factor was deduced by interpolation between
  \fh\ values for other nuclei, which are plotted either as a function
  of the neutron binding energy~\cite{lub94} (see Fig.~\ref{pbhalo})
  or of the asymmetry parameter $\delta =(N-Z)/Z$ ~\cite{jas00}. (Note
  that, as discussed in Ref.~\cite{trz01}, the halo factors published
  by us before this reference should be multiplied by 0.63).  The
  interpolated \fh\ for \nuc{208}{Pb} found in this way is $2.8 \pm
  0.4$.

  This interpolated value can be compared with the results of Bugg
  {\it et al.}~\cite{bug73}, where the idea of the neutron halo factor
  was introduced for the first time to interpret the ratio of charged
  pions generated by antiproton annihilation in various targets,
  including \pb.  In that work the halo factor, defined as $
  f^{\pi}_{halo}= \frac{N(n {\bar p})}{N(p{\bar p})} \frac{Z}{N} \, R$
  (where $N(n {\bar p})$ and $N(p{\bar p})$ are the number of \pbar\
  annihilations on peripheral neutrons and protons, respectively) was
  measured for the \pb\ nucleus by detecting the charged pions emitted
  after antiproton capture in nuclei~\cite{bug73}. The result
  obtained, $ f^{\pi}_{halo}= 2.34(50)$, is based on the assumption
  that $R=0.63$, with this latter number extracted from the \pbar\
  capture in carbon.  The result has been subject to some criticism
  for neglecting the final state interactions and a possible
  dependence of $ R$ on the actual nucleus. The more recent Obelix
  experiments determined $R = 0.48(3)$ from low-energy capture in
  He~\cite{BAL89}, while the best fit optical potential requires $R
  \approx 1$~\cite{fri05}.
  This discrepancy is resolved if one realizes that capture in He
  involves mostly S-waves and capture from high angular momentum
  states in heavier nuclei involves mostly P (or higher) waves in the
  $N\bar{N}$ system~\cite{wyc05}.  We refer to Ref.~\cite{wyc05} for
  some details of the calculation, which estimates $R \approx 0.9 -
  1.0$ in the lead region and for a new analysis of the Bugg result.
  $R=1$ is used in these calculations.  This yields
  $f^{\pi}_{halo}(\mathrm{Pb}) =1.8(4)$.  One can obtain the average
  radius of the absorption $r^{\pi}$ via calculations of the final
  state pion interactions and comparison of the final experimental and
  calculated pion spectra.  This yields $r^{\pi} = c_{ch}+ 1.35$~fm.

  \subsection{\label{analys} Experimental results analyzed by optical 
    potentials}

      \subsubsection{\label{batty} The zero \nn\ range antiproton-nucleus 
        potential of Batty}

      Our previous results~\cite{trz01} for the neutron distribution
      in \nuc{208}{Pb} were obtained using a zero-range \nn\ force
      Batty potential~\cite{bat95,bat97}. This potential, obtained by
      a fit to the antiprotonic X-ray data mainly on light nuclei was
      determined before our results were published. It has, therefore,
      the evident philosophical advantage when compared to the
      recently published potentials strongly~\cite{wyc06} or
      completely~\cite{fri05} relying on our results, including the
      \nuc{208}{Pb} nucleus.

      The neutron-distribution parameters deduced using Batty's
      potential are shown in Table~\ref{2pfpb}.  Calculations were
      done with a code based on the work by Leon~\cite{leo74}.  The
      \dr\ is calculated as the difference between the rms radii of
      the corresponding 2pF point proton (Fricke) and point neutron
      distributions under the assumption of a ``halo type''
      distribution (\dc=0).  The \dr\ value of 0.16~fm differs by
      0.01~fm from the previously published one~\cite{trz01}; this is
      due to the updated value~\cite{codata} of the electromagnetic
      proton rms radius used in the transformation from charge to
      proton densities.

\begin{figure}[h]
\includegraphics[width=0.55\textwidth]{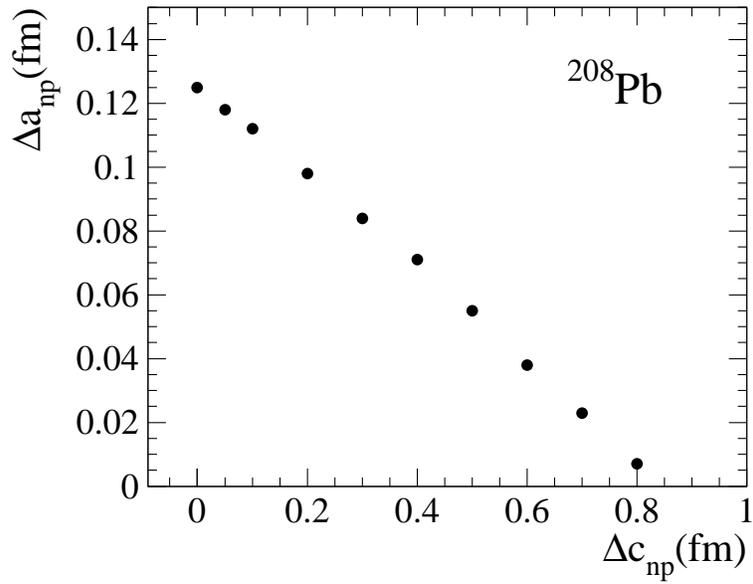}
\caption{\label{dcda} The relation between the difference of the
  diffuseness parameters $a$ and half-density radii $c$ of neutron and
  proton distributions in \nuc{208}{Pb} deduced from the experimental
  antiprotonic level widths using the zero-range interaction Batty
  potential (Ref.~\cite{bat95}).  The experimental uncertainties of
  the level widths are reflected by an almost constant \da\
  uncertainty (not shown) of about $\pm$0.015~fm.}
\end{figure}

\begin{figure}[h]
\includegraphics[width=0.55\textwidth]{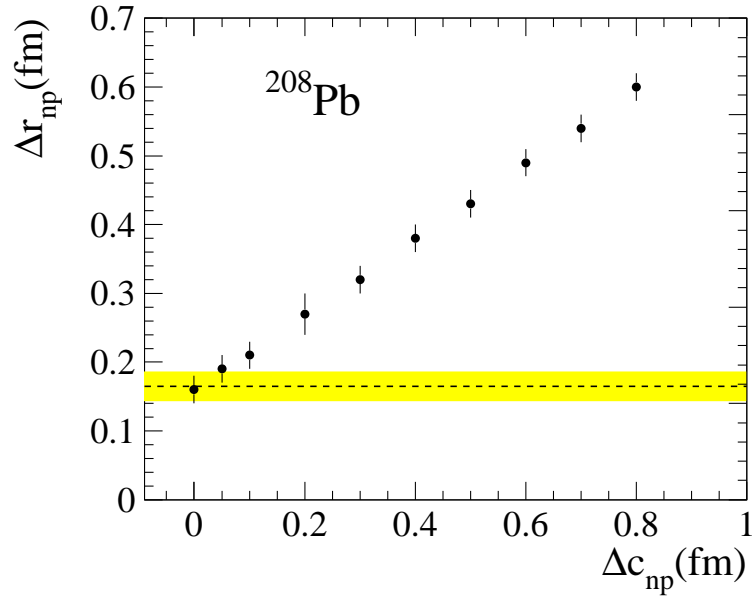}
\caption{\label{drdc} The relation between the difference of the
  neutron and proton half-density radii \dc\ and the deduced
  difference between rms radii of neutron and proton distributions
  \dr\ of \nuc{208}{Pb} for the zero-range interaction Batty
  potential. The dashed line and shaded region shows the weighted
  average and the error, respectively, of \dr\ determined in the
  hadron scattering experiments (see Ref.~\cite{jas04}).  }
\end{figure}

      As indicated above, our experimental data yield $\Delta r_{np} =
      0.16$~fm under the assumption of $c_n = c_p$.  In
      Figs.~\ref{dcda} and~\ref{drdc} we show how the relaxation of
      this condition would influence the difference of the
      Fermi-distribution parameters and the rms radii difference for
      the \nuc{208}{Pb} nucleus.

      The data in Fig.~\ref{dcda} present the change of \da\ when we
      allow the \dc\ to change while still being in agreement with the
      experimental level widths.  It is seen that with the extreme
      neutron-skin assumption (identical proton and neutron
      diffuseness, \da=0) the neutron-proton difference of the half
      density radii \dc\ should be close to 0.8~fm. As shown in
      Fig.~\ref{drdc} such a large value of this difference would lead
      to \dr\ close to 0.6~fm.

      In Ref.~\cite{jas04} we have discussed the results for \dr\ in
      \nuc{208}{Pb} obtained by using hadron scattering data.  The
      weighted average of six experimental results, obtained between
      1979 and 2003, is $\Delta r_{np} = 0.16 \pm
      0.02$~fm~\cite{jas04}.  It is in excellent agreement with the
      result obtained from the antiprotonic X-rays with the zero
      interaction range Batty potential.  The gray band in
      Fig.~\ref{drdc} indicates the error margin of the weighted
      average of the hadron scattering experiments, allowing a
      difference in the \dc\ value between the 2pF distribution of
      neutrons and protons in \nuc{208}{Pb} to be 0.08~fm at most.

      The upper limit for \dc\ can also be estimated comparing the
      calculated neutron to proton density ratio with the interpolated
      halo factor of \pb. This density ratio is plotted in the right
      panel of Fig.~\ref{rorat-halof} as a function of the radius for
      \dr=0.16~fm (Batty potential X-ray value for \dc=0 and average
      value from the hadron scattering data). Two halo factors are
      shown in this figure: one resulting from the pion emission
      experiment and the other one from the interpolation (see
      Sec.~\ref{interpol-halo}). Although the pion experiment is not
      limiting the \dc\ value, the interpolated \fh\ clearly indicates
      that \dc\ has to be smaller than 0.1~fm.  This determines the
      systematic error of the \dr\ value to be equal to 0.04~fm. If
      Ref.~\cite{jag74} instead of Ref.~\cite{fri95} is taken for the
      charge distribution, \dr=0.12~fm is obtained, i.e. the
      systematic error is also 0.04~fm for this case (cf. also results
      of Ref.~\cite{wyc06} in Sec.~\ref{wycech}).

      The assigned statistical and systematic error for the \dr\ value
      indicates about 1\% uncertainty in the determination of the
      neutron rms radius in \pb\ from the antiprotonic atom data. This
      value is comparable to the expected precision of the parity
      violation measurements~\cite{hor01c} of this quantity.

      The comparison of the experimentally determined level widths and
      shift with the theoretical proton and neutron distributions (see
      Sec.~\ref{calc-mean}) using Batty's potential is shown in
      Table~\ref{gamcalc}.  The level widths calculated with the SkP
      and SkX distributions are too small (by 12\% and 27\%,
      respectively) whereas the DD-ME2 distributions result in widths
      close to the experimental values.  As indicated in
      Sec~\ref{potential}, the value of the shift is not reproduced
      for any theoretical distribution.

      It is interesting to note that the SkX interaction with the
      value closest to the experimental \dr\ leads to level widths
      which are clearly below the observed ones. It indicates that for
      this interaction the nucleon density decreases too fast with
      $r$.  This was illustrated in Fig.~\ref{nplusp} by a comparison
      of the summed neutron and proton densities for the three forces
      considered.  As previously discussed in~\cite{kar02}, too small
      proton but especially neutron diffuseness exhibited by the SkX
      model is the reason for this behavior
      (cf. Table~\ref{2pfpb}). Our antiprotonic atom data presented in
      Table~\ref{gamcalc} are therefore a confirmation of the
      conclusions drawn from the analysis of nucleon elastic
      scattering, presented in Ref.~\cite{kar02}.

      \subsubsection{\label{friedman} The finite \nn\ range potential
       of Friedman}
     
      Contrary to the Batty potential~\cite{bat95,bat97}, the
      finite-range antiproton-nucleus interaction potential recently
      proposed by Friedman {\it et al.}~\cite{fri05} was based almost
      completely on the PS209 experimental data, including the
      antiprotonic atom level widths and shifts of \pb\ and \bi\
      reported in this publication.

\begin{figure}[h]
\includegraphics[width=0.55\textwidth]{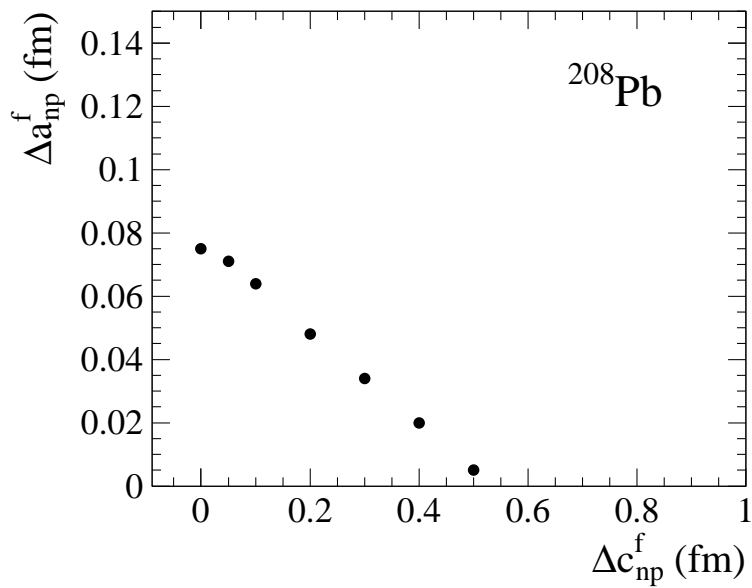}
\caption{\label{dcdaf} The same as Fig.~\ref{dcda}, but for 
  the finite-range Friedman potential~\cite{fri05}.}
\end{figure}

\begin{figure}[h]
\includegraphics[width=0.55\textwidth]{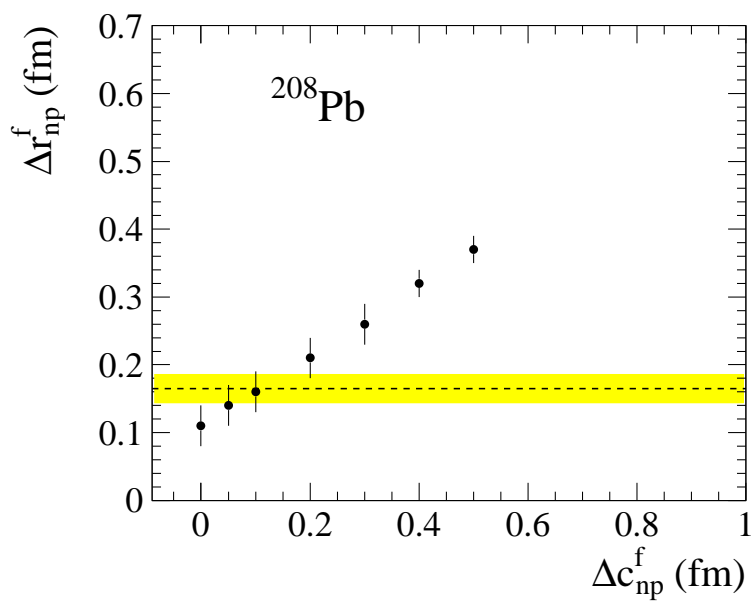}
\caption{\label{drdcf} The same as Fig.~\ref{drdc}, but for 
  the finite-range Friedman potential~\cite{fri05}.}
\end{figure}

      In order to obtain a similar relationship as that shown for
      Batty's potential in Fig.~\ref{dcda}, the 2pF distribution of
      protons (see Table~\ref{2pfpb}) was folded over the interaction
      range with 1.04~fm rms radius. The (folded) neutron density was
      fitted using the optical potential parameters from~\cite{fri05}
      to the experimental level widths of \pb, varying \dcf. The \daf\
      values of the folded densities were deduced from the fit and are
      shown in Fig.~\ref{dcdaf}. Figure~\ref{drdcf}, similar to
      Fig.~\ref{drdc}, shows the resulting \drf\ values of the folded
      density distributions as a function of the folded \dcf\ values.

      It can be shown that the transformation from the point-like
      nucleon distributions to the folded ones increases the \daf\
      values by about 15\%, leaving \dc\ and \dr\ approximately
      unchanged.  Therefore, without any supplementary conditions
      these results indicate that using the finite-range version of
      the optical potential as proposed by~\cite{fri05}, the
      \drf$\approx$\dr\ value obtained for \pb\ is between 0.11~fm
      (\dcf=0~fm) and 0.38~fm (\dcf=0.54~fm).
     
      In Ref.~\cite{fri05} it was shown that if all antiprotonic data
      are presented in the form $\Delta r_{np} = \alpha (N-Z)/A +
      \beta$~\cite{trz01} the global fit to these data allows $\alpha$
      values between 0.9~fm and 1.3~fm (our previous data
      in~\cite{trz04} analyzed in terms of the point proton optical
      potential gave $\alpha=0.90 \pm 0.15$~fm). This $\alpha$ range
      gives \dr\ values between 0.154~fm and 0.240~fm.
     
      Another limit of the \dr\ values could be obtained with the help
      of the weighted average of the hadron scattering data, giving
      \dr=0.16$\pm$0.02~fm~\cite{jas04}. Two standard deviations of
      this average (cf.  Fig.~\ref{drdcf}) put the lower and upper
      limit of the \dr\ as 0.12~fm and 0.20~fm.

      Taking the lower allowed value from the global fit and the upper
      one from the hadron scattering gives \drf$\approx$\dr=0.17~fm as
      the result of the finite range Friedman potential with the
      estimated error of $\pm$0.02~fm.

      This result implies \dcf$\approx$\dc\ of 0.13~fm, i.e. the
      neutron distribution essentially of the ``halo type'' but with a
      small contribution of the ``skin type''. Such a distribution is
      shown in Fig.~\ref{rorat-halof} by the curve labeled ``G'',
      which lies slightly below the lower limit of the interpolated
      halo factor.

      The comparison of the experimentally determined level widths and
      shift with the theoretical proton and neutron distributions
      using the Friedman potential is given in the lowest part of
      Table~\ref{gamcalc}. The calculated level widths have about 5\%
      uncertainty due to the folding procedure applied. Within these
      errors calculated level widths are identical to those obtained
      using Batty's potential.

      \subsubsection{\label{wycech} A constraint finite \nn\ range potential}

      A parallel study with the potential from Ref.~\cite{wyc06}
      generates two solutions \dr=0.16(3)~fm for the electron
      scattering charge~\cite{jag74} and/or \dr=0.22(3)~fm for the
      muonic charge density~\cite{fri95}.  These solutions favor halo
      type neutron densities, both are characterized by large $\chi^2$
      values due to the poorly reproduced level shift.

    \section{Analysis and discussion of  \nuc{209}{Bi} data}

    Since the \nuc{209}{Bi} nucleus has the spin $I = 9/2$ and
    magnetic moment $\mu = 4.08 \cdot \mu_N$ (where $\mu_N$ is the
    nuclear magneton), the antiprotonic atom levels are split. The
    related hyperfine shifts are smaller than the fine structure
    (f.s.) splitting but comparable to the strong interaction
    shifts. The standard formula~\cite{BET}, extended to the case of
    an anomalous magnetic moment~\cite{BAR81}, gives
    \begin{equation}
      \label{label1}
      E_{hfs} = \frac{\alpha  g_{\bar{p}}g_{Bi}}{8mM}
      \frac{ [F(F+1)- J(J+1)- I(I+1)] }{ J(J+1)n^3(l+1/2)}(m Z \alpha  )^3,
    \end{equation}
    where $M$ is the proton mass, $m$ is the reduced mass, $ g I = \mu
    $ in nuclear magnetons.  The last transition in Bi is split into
    10 dominant components in the upper f.s. state and into 9
    components in the lower f.s.  components. These correspond to
    different values of the total spin of the system $F= J +I,...,J-I
    $.  Assuming a statistical population $\sim(2F +1)$, the observed
    spectral line becomes asymmetrical.  One obtains an overall 50~eV
    repulsive shift of the centroid and an additional "broadening" of
    150~eV.  This yields a lower width of 350(50)~eV and an attractive
    lower shift of 37(53)~eV, generated by strong interaction. (The
    upper level width averaged over the fine structure components (see
    Table~\ref{gamut}) is equal to 6.9(1.3)~eV). One thus faces a
    sizable isotopic effect between attraction in $^{209}$Bi and
    repulsion in $^{208}$ Pb atoms. The difference is related to the
    weakly bound valence proton in this Bi isotope. As discussed on
    previous occasions, there are indications of a quasi-bound state
    in the p\pbar\ system just below the threshold. Such a state
    generates few distinct cases of an anomalous behavior of level
    shifts in nuclei with loosely bound nucleons~\cite{wyc01}.  For a
    more quantitative discussion of these phenomena we refer to a
    parallel publication~\cite{wyc06}.
  
    Assuming (as for \pb) \dc=0~fm, the difference of the neutron and
    proton diffuseness parameter was fitted to level widths using
    Batty's potential.  The \dr\ value obtained is equal to
    0.14$\pm$0.04~fm, lower by 0.04~fm than the previously published
    one~\cite{trz01}. In Ref.~\cite{trz01} the hyperfine splitting of
    the lower level discussed above was not taken into account.

    \section{Summary and conclusions}

    In this article we have presented an analysis of the nuclear
    structure information extracted from the studies of the
    antiprotonic atoms \pbar-\pb\ and \pbar-\bi.  The experimentally
    determined level widths and shifts of these atoms at the end of
    the antiprotonic cascade depend on the antiproton-nucleus
    interaction potential. In turn, the crucial ingredient of this
    potential is the nucleon density at the radial distance where the
    antiproton-nucleus interaction occurs.  Therefore, as it was shown
    already in our first publications in this
    series~\cite{jas93,lub94}, the study of antiprotonic atoms may
    constitute a powerful tool for the extraction of information on
    the properties of the nuclear periphery.

    In the analysis of the antiprotonic atom data presented here we
    have been essentially using two optical antiproton-nucleus
    potentials, proposed by Batty {\it et al.}~\cite{bat95,bat97} and
    Friedman {\it et al.}~\cite{fri05}, respectively.  The Batty
    potential, now more than ten years old, was obtained by fitting
    the potential parameters to the 33 level widths and 15 level
    shifts of antiprotonic atoms published at that time, mainly of
    light and a few intermediate-mass nuclei.  The fits were performed
    with a zero-range antiproton-nucleon interaction.  Although this
    unphysical assumption is presently avoided~\cite{fri05,wyc06} as
    leading to worse fits than the finite range potentials, we still
    pursue our data analysis using the Batty prescription in its
    simplest form. Our arguments are that although ``point nucleon
    distributions'' and ``zero range interaction potentials'' are
    probably an oversimplification, the obtained parameters are
    deduced from the fit to the experimental data.  It may be expected
    that this fact somehow in an automatic way introduces the
    corrections of the method deficiencies. Moreover, as already
    mentioned at the beginning of Sec.~\ref{batty}, the
    Batty-potential parameters were obtained before our antiprotonic
    atom data were available, ensuring the interpretation of the
    results to be independent of the interpretation tools.

    Recently, an antiproton-nucleus optical potential with finite
    interaction range was proposed by Friedman \ea~\cite{fri05}. It
    was shown that the 90 data points from our PS209 X-ray
    experiments, together with 17 data points from the radiochemical
    experiment, determine an attractive and absorptive \pbar-nuclear
    isoscalar potential, which fits the data well.

    However, as the \pb\ antiprotonic X-ray data were used in the
    determination of the Friedman potential, we have tried to show in
    the analysis of the experiment with this potential what can be
    deduced on the neutron distribution of this nucleus on a more
    general ground.  To this end we have used the information on the
    trend of the \dr\ values as a function of the asymmetry parameters
    $(N-Z)/A$, allowed by the potential of Ref.~\cite{fri05}, and
    previously analyzed~\cite{jas04} results of the hadron scattering
    experiments.

    Another fit with a finite-range potential was recently proposed
    (Ref.~\cite{wyc06}, see also Sec.~\ref{wycech}). The reader is
    referred to the original reference for the discussion of the
    global fit to the antiprotonic X-ray data performed and the \dr\
    values deduced for \pb.  These values are used in the present
    publication to estimate the systematic errors.

    It was shown in this paper that by an analysis based strictly on
    the experimental antiprotonic level widths one would be unable to
    propose meaningful limits for the \dr\ value in \pb. Applying the
    Batty potential with a pure ``halo shape'' of the neutron
    distribution (\dc=0~fm) leads to \dr=0.16~fm. The lowest limit of
    the interpolated halo factor would allow at most \dc=0.1~fm,
    i.e. \dr=0.20~fm, similarly to the $2\sigma$ uncertainty of the
    average \dr\ value deduced from the hadron scattering
    experiments. The analyzed theoretical proton and neutron
    distributions using HF, HFB and RMF models give a \dc\ value of
    0.07~fm at most.  If the charge distribution from
    Ref.~\cite{jag74} is used instead of that from Ref.~\cite{fri95},
    \dr=0.12~fm is obtained.  We conclude that our experiment
    interpreted using Batty's potential with the supplementary
    information given above leads to \dr=$0.16 \pm (0.02)_{stat} \pm
    (0.04)_{syst}$~fm in the \pb\ nucleus.  A value for \dr\ that is
    only 0.01~fm larger is deduced from the analysis using the
    Friedman potential.

    Significant results were obtained applying the Batty and Friedman
    potentials with the theoretical proton and neutron distributions
    to get the antiprotonic \pb\ atom level widths and shift. As it
    was discussed in Sec.~\ref{potential} the experimental level shift
    was not reproduced.  On the other hand for each theoretical
    distribution both potentials give almost identical widths in spite
    of the fact that one is ``zero range'' and the other one is
    ``finite range'' interaction potential.  The calculated widths are
    smaller in some cases than the experimental ones.  This is
    interpreted as evidence obtained from antiprotonic atoms for too
    rapid a decrease of these theoretical nucleon densities as a
    function of the radial distance due to a too small diffuseness of
    these densities. A similar conclusion was previously obtained from
    the analysis of nucleon elastic scattering.

    \begin{acknowledgments}
      We are grateful to Prof. E.~Friedman for illuminating
      discussions.  Financial support by the Polish State Committee for
      Scientific Research as well as by Deutsche
      Forschungsgemeinschaft Bonn (436POL17/8/04) is acknowledged.
    \end{acknowledgments}

\clearpage

{\bf\large Tables}

\begin{table}[h]
\caption{\label{target} Target properties and number of antiprotons used.}
\begin{ruledtabular}
\begin{tabular}{c|ccc}
Target &  thickness d  & enrichment & number of $\bar p$ (10$^8$)  \\ 
  &  (mg/cm$^2$) & (\%) &   \\
\hline
$^{208}$Pb & 130.4 & 99.1 & 17 \\
$^{209}$Bi & 132.7&  nat. & 1.4 \\
\end{tabular}
\end{ruledtabular}
\end{table}

\newpage

\begin{center}
\begin{longtable}{c|cc}
\caption{\label{relat} Measured relative  X-ray intensities, normalized to the 
transition n\,=\,13 $\to$ 12 (average value from three detectors).} \\ \hline
Transition & $^{208}$Pb  & $^{209}$Bi \\ \hline
\endhead
\multicolumn{3}{r}{Continued on next page} \\ \hline
\endfoot
\hline \hline
\endlastfoot
10$ \to$ 9  &  69.0  $\pm $  2.3  &  63.6  $\pm $   3.3   \\
11$ \to$ 10 &  99.6  $\pm $  3.2  &  97.8  $\pm $   3.6   \\
12$ \to$ 11 &  103.8 $\pm $  4.1  &  102.1 $\pm $   4.7   \\
13$ \to$ 12 &  100.0  $\pm $  4.6  & 100.0  $\pm $   4.9  \\
14$ \to$ 13 &  93.5  $\pm $  5.0  &  93.8  $\pm $   5.4   \\
15$ \to$ 14 &  81.3  $\pm $  3.7  &  81.2  $\pm $   4.0   \\
16$ \to$ 15 &  59.4  $\pm $  2.7  &  61.4  $\pm $   3.1   \\
17$ \to$ 16 &  54.8  $\pm $ 10.9$^*$  &  80.3  $\pm $   10.4$^*$ \\
18$ \to$ 17 &  73.4  $\pm $  3.6  &  71.2  $\pm $   4.0      \\
19$ \to$ 18 &  56.0  $\pm $  2.6  &  56.6  $\pm $   2.9      \\
20$ \to$ 19 &  48.2  $\pm $  2.3  &  53.4  $\pm $   4.8      \\
21$ \to$ 20 &  44.9  $\pm $  3.3  &  67.8  $\pm $   6.3      \\
 & &  \\
11$ \to$ 9  &  3.3   $\pm $  0.3 &                   \\
12$ \to$ 10 & 7.6 $\pm$ 0.5  & 5.3  $\pm $   0.8 \\
13$ \to$ 11 &  10.1  $\pm $  0.3 & 9.1 $\pm $   0.5 \\
14$ \to$ 12 &  11.2  $\pm $  0.6 & 10.4 $\pm $   0.8 \\
15$ \to$ 13 &  10.6  $\pm $  0.5 & 11.2 $\pm $   0.8   \\
16$ \to$ 14 &  10.5  $\pm $  0.5 & 10.7 $\pm $   0.8  \\
17$ \to$ 15 &  12.2  $\pm $  0.6 & 11.4 $\pm $   0.8  \\
18$ \to$ 16 &  12.0  $\pm $  0.4 & 11.2 $\pm $   0.6  \\
19$ \to$ 17 &  11.8  $\pm $  0.6 & 11.0 $\pm $   0.8  \\
20$ \to$ 18 &  10.1   $\pm $  0.5 & 11.0 $\pm $   0.8  \\
21$ \to$ 19 &  5.2   $\pm $  2.8$^*$ &      1.3 $\pm$ 1.0 $^*$ \\
22$ \to$ 20 &     0.0 $\pm$ 7.9$^*$    &      0.0 $\pm$ 7.8$^*$  \\
23$ \to$ 21 &  4.2   $\pm $  0.2 & 9.6  $\pm $  1.1   \\
24$ \to$ 22 &  3.5   $\pm $  0.3 & 4.4  $\pm $  0.6    \\
25$ \to$ 23 &  5.0    $\pm $  0.3 & 5.1  $\pm $  0.7  \\
            &                     &   \\
14$ \to$ 11 & 3.1   $\pm $  0.3   &     \\
15$ \to$ 12 & 2.6   $\pm $  0.2   & 3.3  $\pm $  1.1   \\
16$ \to$ 13 & 2.4   $\pm $  0.2   & 3.0  $\pm $  0.6     \\
17$ \to$ 14 & 4.2   $\pm $  0.3   & 2.4  $\pm $  0.5    \\
18$ \to$ 15 & 3.7   $\pm $  0.2   & 3.1  $\pm $  0.5    \\
19$ \to$ 16 & 3.5   $\pm $  0.2   & 2.8  $\pm $  0.5    \\
20$ \to$ 17 & 3.5   $\pm $  0.2   & 2.8  $\pm $  0.4    \\
21$ \to$ 18 & 3.7   $\pm $  0.2   & 2.9  $\pm $  0.3     \\
22$ \to$ 19 & 3.9   $\pm $  0.2   & 3.0  $\pm $  0.4     \\
23$ \to$ 20 & 3.3   $\pm $  0.2   & 2.4  $\pm $  0.5    \\
24$ \to$ 21 & 1.4   $\pm $  0.3   & 3.4  $\pm $  0.7     \\
25$ \to$ 22 & 3.0   $\pm $  0.2   & 2.0  $\pm $  0.4      \\
26$ \to$ 23 & 10.3   $\pm $  0.6   & 11.5 $\pm $  0.8    \\
27$ \to$ 24 & 9.8   $\pm $  0.5   & 3.5  $\pm $  0.7     \\
28$ \to$ 25 & 3.5   $\pm $  0.3   &              \\
                                       &      &                    \\
17$ \to$ 13 & 3.3   $\pm $  0.3   &  \\
18$ \to$ 14 & 1.2   $\pm $  0.2   &  \\
19$ \to$ 15 & 1.6   $\pm $  0.2   &  \\
20$ \to$ 16 & 1.7   $\pm $  0.2   &   \\
21$ \to$ 17 & 2.1   $\pm $  0.2   &  \\
22$ \to$ 18 & 2.8   $\pm $  0.2   &  \\
23$ \to$ 19 & 1.2   $\pm $  0.1   &  \\
24$ \to$ 20 & 1.3   $\pm $  0.2   &  \\
25$ \to$ 21 & 1.1   $\pm $  0.2   &  \\
\multicolumn{3}{l}
{$^{*}$ admixtures of electronic X-rays from the same atom} \\
\multicolumn{3}{l}
{and from the (Z$\pm$1) atoms were subtracted.}\\
\end{longtable}
\end{center}

\clearpage
\newpage

\begin{table}[h]
\caption{\label{shiftt} Measured shifts of the $(n,l=10,9)$ ($\epsilon_u$) 
and $(n,l=9,8)$ ($\epsilon_l$) levels in 
the antiprotonic \nuc{208}{Pb} and \nuc{209}{Bi} atoms.}
\begin{ruledtabular}
\begin{tabular}{ccccc}
Target & $\varepsilon_u^+$\,(eV) &   $\varepsilon_u^-$\,(eV)  & $\varepsilon_l^+$\,(eV) & $\varepsilon_l^- $\,(eV)      \\
\hline
$^{208}$Pb  & 34 $\pm$ 16 & 28 $\pm$ 17  & 102 $\pm$ 28 & 73 $\pm$ 29 \\
$^{209}$Bi  & 23 $\pm$ 20 & -8 $\pm$ 21  &  29 $\pm$ 72 & -2 $\pm$ 73   \\
\end{tabular}
\end{ruledtabular}
\end{table}

\begin{table}[h]
\caption{\label{gamlt} Measured absorption widths of the 
fine structure components of the $(n,l=9,8)$ 
level in the antiprotonic
\nuc{208}{Pb} and \nuc{209}{Bi} atoms.}
\begin{ruledtabular}
\begin{tabular}{ccc}
Target& $\Gamma_l^+$\,(eV) & $\Gamma_l^-$\,(eV) \\
\hline
$^{208}$Pb & 320 $\pm$ 35 & 302 $\pm$ 38    \\
$^{209}$Bi  & 557 $\pm$ 68 & 448 $\pm$ 74 \\
\end{tabular}
\end{ruledtabular}
\end{table}

\begin{table}[h]
\caption{\label{gamut} Radiation width $\Gamma _{em}$ and Auger width 
$\Gamma _{Auger}$ for the n=10 levels, where the strong interaction width 
$\Gamma_u$ was determined via the intensity balance.} 
\begin{ruledtabular}
\begin{tabular}{ccccc}
Target  & $\Gamma _{em}$\,(eV) & $\Gamma _{Auger}$\,(eV) & $\Gamma_u^+$\,(eV) & $\Gamma_u^-$\,(eV)  \\
\hline
$^{208}$Pb  &  12.59 & 0.139 &  5.3 $\pm$ 1.1 & 6.6 $\pm$ 1.3  \\
$^{209}$Bi  & 13.27 & 0.141  & 6.1 $\pm$ 1.7 & 7.8 $\pm$ 1.9  \\
\end{tabular}
\end{ruledtabular}
\end{table}

\begin{table}
\caption{ \label{2pfpb} Comparison of 2pF proton and neutron
  distributions in $^{208}$Pb (all parameters in fm).}
\begin{ruledtabular}
 \begin{tabular}{|l||l|l|l|l||l|l|l|l||l|l|l|l|}
 & \multicolumn{4}{c||}{protons} & \multicolumn{4}{c||}{neutrons}  &  \multicolumn{4}{c|}{} \\
\hline
 & \multicolumn{2}{c|}{\rp} &\app &\cp  & \multicolumn{2}{c|}{\rn} &\an      &\cn & \da  & \dc & \multicolumn{2}{c|}{\dr}\\
\hline
 & $^{(a)}$ & $^{(b)}$ & & & $^{(a)}$ & $^{(b)}$ & & & & &$^{(a)}$  & $^{(b)}$  \\  
SkP   & 5.465 & 5.489 & 0.437 & 6.768 &  5.610 & 5.625 & 0.537 & 6.789 & 0.100 & 0.021 & 0.145 & 0.136 \\
SkX   & 5.441 & 5.443 & 0.424 & 6.726 &  5.597 & 5.597 & 0.510 & 6.799 & 0.086 & 0.073 & 0.156 & 0.154 \\ 
DD-ME2& 5.460 & 5.472 & 0.444 & 6.736 &  5.653 & 5.657 & 0.561 & 6.789 & 0.117 & 0.053 & 0.193 & 0.185 \\
\hline
Fricke $^{(c)}$ 
       &      & 5.436 & 0.446 & 6.684 &        &       &       &       &       &       &       &      \\ \hline
Experiment$^{(d)}$      
       &      &       &       &       &        & 5.596 & 0.571 & 6.684 & 0.125 & 0.0$^{(e)}$ & & 0.16(2) \\ 
\end{tabular}
\end{ruledtabular}
\begin{tabular}{l}
{$^{(a)}$ calculated from the theoretical distributions;}\\
{$^{(b)}$ calculated from fit parameters: \app ,\cp , \an , \cn;}\\
$^{(c)}$ point proton values obtain from the Fricke \cite{fri95} charge
 distribution using Oset's \cite{ose90} transformation\\
 formulae; \\
$^{(d)}$ from 2pF fit to the experimental width using the Batty 
potential~\cite{bat95}; \\
$^{(e)}$ assumed. 
\end{tabular}
\end{table}

\begin{table}
\caption{\label{annihil-probab-tab} Parameters of the annihilation
  probability distribution calculated using DD-ME2 density and
  the zero-range Batty potential. }
\begin{ruledtabular}
\begin{tabular}{cc|cccc}
             &       &\multicolumn{4}{c}{radial parameter (fm)} \\
Distribution & Level & FWHM & most probable & median & average \\ \hline
DD-ME2       & up    & 1.5 &   8.3         &  8.7   &  9.1    \\
             & low   & 1.2 &   8.5         &  8.6   &  8.8    \\ 
\end{tabular}
\end{ruledtabular}
\end{table}

\begin{table}
\caption{\label{gamcalc} Comparison of experimental level widths and shift in \pb\
with those calculated
using theoretical neutron and proton distributions. }
\begin{ruledtabular}
\begin{tabular}{cccc}
      & \gl  &  \gu  & $\epsilon$ \\
      & eV   &   eV  &  eV  \\ \hline
Experiment & 312(26) & 5.9(8) & 88(20) \\ \hline
Batty potential & & &  \\
SkP   & 274  & 5.2   & 14 \\
SkX   & 231  & 4.2   & 16 \\
DD-ME2& 315  & 6.2   & 12 \\ \hline
Friedman potential & & & \\ 
SkP   & 278  &  5.3  & 6 \\
SkX   & 244  &  4.5  & 7 \\
DD-ME2& 307  &  6.1  & 2 \\
\end{tabular}
\end{ruledtabular}
\end{table}

\clearpage

{\bf \large Figures}

\clearpage

\end{document}